\documentclass[11pt]{article}

\usepackage{epsfig}

\newcommand{\be}{\begin{equation}}
\newcommand{\ee}{\end{equation}}
\newcommand{\bea}{\begin{eqnarray}}
\newcommand{\eea}{\end{eqnarray}}
\newcommand{\bra}{\langle}
\newcommand{\ket}{\rangle}

\begin{document}

%\markboth{Authors' Names}
%{Instructions for Typing Manuscripts (Paper's Title)}

%\catchline{}{}{}{}{}
%
%%%%%%%%%%%%%%%%%%%%%%%%%%%%%%%%%%%%%%%%%%%%%%%%%%%%%%%%%%%%%%%%%%%%

\title{Density of States Method at Finite Isospin Density}

\author{Tetsuya TAKAISHI \\
%\footnote{
%Typeset names in 8-pt Times Roman, uppercase. Use the footnote to 
%indicate the present or permanent address of the author.}
%\address{
Hiroshima University of Economics,\\
Hiroshima, 731-0192,
Japan}
%\footnote{
%Hiroshima University of Economics,
%5-37-1, Gion, Asaminami-ku, Hiroshima 731-0192
%%E-mail: takaishi@hiroshima-u.ac.jp
%}\\
%takaishi@hiroshima-u.ac.jp}

\date{}
\maketitle

%\pub{Received (Day Month Year)}{Revised (Day Month Year)}

\begin{abstract}
The density of states method is applied
for lattice QCD at a finite isospin density.
The advantage of this method is that
one can easily obtain results for various values of parameters ( quark mass,
coupling constant and the number of flavors ).
We compare results for the chiral condensate and the quark number density
with  those from the R-algorithm and find that they are in good agreement.
By calculating the chiral condensate we obtain information on the phase structure for
various quark flavors and isospin chemical potentials.
We also show results for the chiral condensate at two different quark masses and
at two different isospin densities which are not easily obtainable in
the conventional Monte Carlo method.

%\keywords{Lattice QCD calculations; Density of states method; Finite density}
\end{abstract}

%\ccode{PACS Nos.:11.15.Ha, 12.38.Gc
%}

\section{Introduction}
The lattice QCD Monte Carlo technique has been a useful tool 
for clarifying the non-perturbative aspects of QCD 
at the zero baryon chemical potential ( $\mu_B=0$ ) 
for both finite and zero temperatures.
At nonzero $\mu_B$, however, due to the sign problem,
the standard importance sampling method fails.
The Glasgow group\cite{Glasgow} attempted to extract information on $\mu_B$ at low temperature by 
a method based on  reweighting but this method does not work at low temperatures
due to the overlap problem.
Recently it has been realized that at low density and finite temperature 
the sign problem may not be a serious numerical difficulty   
and a variety of approaches ( multi-reweighting, Taylor expansion, imaginary chemical potential)
\cite{RECENT1,RECENT2,RECENT3,RECENT4,RECENT5}
have been applied to study QCD at low density and finite temperature 
(See \cite{REVIEW1,REVIEW2,REVIEW3,REVIEW4}for  recent reviews).
For larger $\mu_B$ we are still lacking an efficient numerical method. 

In contrast to the finite $\mu_B$ case, QCD at isospin density has no sign problem and 
can be simulated by lattice Monte Carlo methods\cite{ISOSPIN}.  
There is an expectation that at small isospin density the system 
may resemble that at small baryon density\cite{RECENT5,Kogut}. 
Recent studies have shown that the phase diagram at small isospin density is
similar to that at small $\mu_B$\cite{Kogut2,AT}.
On the other hand, a difference between the isospin and baryon density
is seen in susceptibilities of number density and derivatives of meson mass\cite{RECENT1,SUS}.   

In lattice Monte Carlo simulations, usually 
importance sampling is used.
This approach have been proven to be very efficient 
by many studies. 
Another approach one may take is 
the density of states (DOS) method.
The difference between importance sampling and 
the DOS method is as follows.
In importance sampling a simulation is performed 
at a fixed parameter set. To explore different parameters 
one must perform independent simulations\footnote{If the parameter region to be explored is 
narrow, then one may apply the reweighting method\cite{SWENDSEN} which does not need independent simulations.}. 
On the other hand, in the DOS method 
one first determines the DOS of suitable observables 
and then results  are obtained by performing one or a few dimensional integral.    
To illustrate how the DOS method works, let us consider a gauge model. 
The partition function for this model is given by
$
\displaystyle
Z=\int [dU] \exp(-\beta S_g[U]),
$
where $S_g[U]$ is a gauge action and $\beta$ is a coupling constant.
If we define the DOS by 
$
\displaystyle
n(E)=\int [dU] \delta(E-S_g[U]),
$
then the partition function is rewritten as
$
\displaystyle
Z=\int dE n(E)\exp(-\beta E).
$
If we consider the average value of the gauge action,
it is given by
$<E>=\int dE n(E)E\exp(-\beta E)/Z$.
This is a one-dimensional integral of $E$ and 
once we obtain $n(E)$, we can evaluate this easily at $any$ $\beta$.
No extra simulation is needed to calculate it at various $\beta$,
which is considered to be an advantage of this method.

The DOS method has been applied for gauge theories:
Z(n), SU(2) and SU(3) models\cite{GAUGE}.
Although the inclusion of dynamical fermions is computationally difficult,
in Ref.\cite{Azcoiti} the DOS method was applied for QED,
where the method is called the microcanonical fermionic average method.
Recently, Luo\cite{LUO} extended the idea of the DOS method to QCD and emphasized that
once the eigenvalues of the Dirac operator and the DOS of the plaquette energy
are determined, 
one can evaluate the thermodynamic quantities derived from the partition function
at $any$ quark mass and flavor.

In this study we apply the DOS method for QCD at isospin density on a $4^4$ lattice
and demonstrate how the DOS method works.
Since the isospin system has no sign problem,
results from the DOS method can be compared with those from the standard method. 
In Sec.2 we give general formulas for forming 
the DOS including dynamical fermions.
In Sec.3 we give simulation details. 
Results are presented in Sec.4.
We summarize our results in Sec.5.

\section{Density of States Method}

\subsection{General Formulas}
In lattice QCD Monte Carlo simulations, usually we aim at 
obtaining the average values of observables:
\be
\bra O \ket =\frac1{Z}\int [dU] O[U] \det \Delta(m_q,\mu)^{N_f/4} \exp(-\beta S_g[U]),
\label{IMPORTANCE}
\ee
where $N_f$ is the number of flavors and
$\Delta(m_q,\mu)$ is assumed to be a staggered fermion matrix at quark mass $m_q$ 
and at chemical potential $\mu$: 
\bea
\Delta(m_q,\mu)_{i,j} =  m_q\delta_{i,j}
                     & + & \frac12\sum_{\nu=\hat{1},\hat{2},\hat{3}}\eta_{i,\nu}(U_{i,\nu}\delta_{i,j-\nu}
-U^\dagger_{i-\nu,\nu}\delta_{i,j+\nu})  \\ \nonumber
&  + &\frac12\eta_{i,\hat{4}}(e^\mu U_{i,\hat{4}}\delta_{i,j-\hat{4}}
-e^{-\mu} U^\dagger_{i-\hat{4},\hat{4}}\delta_{i,j+\hat{4}}).
\eea
$S_g[U]$ is the standard Wilson gauge action,
\be
S_g[U]=\frac1{N_c}\sum_p Re Tr(U_p),
\ee
where $U_p$ is the plaquette and $N_c=3$ for $SU(3)$ gauge theory.
For Wilson fermions, $N_f/4$ in eq.(\ref{IMPORTANCE}) should be replaced with $N_f$.
The standard means of dealing with this multi-dimensional integral is 
to do the importance sampling, i.e. configurations are generated 
with a measure $\displaystyle \sim dU \det \Delta^{N_f/4} e^{-\beta S_g}$
and $\bra O\ket$ is given by an average over the configurations 
as $\bra O\ket \approx \frac1N \sum^N_{i=1} O[U_i]$, where $N$ is 
the number of configurations.

The DOS method rewrites the partition function and 
reduces it to a few dimensional integral.
Let us define the DOS with parameters $E_i$ ($i=1,\dots,k$) as
\be
n(E_1,E_2,\dots,E_k) =\int [dU] g(U) \Pi_i \delta(E_i-x_i(U)),
\label{GDOS1}
\ee
where $x_i(U)$ is an operator associated with $E_i$ and
$g(U)$ is introduced to generalize the DOS further with $g(U)\neq 1$.
Using the DOS, we obtain

\be 
\bra O\ket = \frac1{Z_n}\int [\Pi_i dE_i] n(E_1,E_2,\dots,E_k)
\bra O\det\Delta^{N_f/4} e^{-\beta S_g}/g(U)\ket_{E_1,E_1,\dots,E_k},
\ee
where 
\be
Z_n=\int [\Pi_i dE_i] n(E_1,E_2,\dots,E_k)\bra \det \Delta^{N_f/4} 
e^{-\beta S_g}/g(U)\ket_{E_1,E_2,\dots,E_k}.
\ee
Here, $\bra A\ket_{E_1,E_2,\dots,E_k}$ stands for  an average of $A$ 
on configurations generated with the measure $ [dU] g(U) \Pi_i \delta(E_i-x_i(U))$,
or one can write it as
\be
\bra A \ket_{E_1,E_2,\dots,E_k} =\frac1{n(E_1,E_2,\dots,E_k)}\int dU g(U)\Pi_i \delta(E_i-x_i(U))A(U) .
\ee
For $g(U)=1$, $\bra A\ket_{E_1,E_2,\dots,E_k}$ becomes a microcanonical average at fixed $E_1,E_2,\dots,E_k$.

\subsection{$g(U)=1$ }
Here we give the formulas used in the present study.
We consider the  case of one parameter  and
choose $x(U)=S_g[U]$. 
Although there are many possibilities to choose $x(U)$, 
this choice with $g(U)=1$ is the one used in Ref.\cite{LUO} and is useful for our purpose.
Setting $g(U)=1$, 
we obtain 
\be
n(E) =\int [dU] \delta(6VE-S_g[U]),
\label{dosn}
\ee
\be
\bra O\ket = \frac1{Z_n}\int  dE n(E) e^{-\beta 6VE}
\bra O\det\Delta(\mu)^{N_f/4} \ket_{E},
\label{dosn2}
\ee
and 
\be
Z_n=\int  dE n(E)e^{-\beta 6VE}\bra \det \Delta(\mu)^{N_f/4}
\ket_{E},
\label{dosn3}
\ee
where $V$ is the number of lattice sites and $E$ is the plaquette energy.
These are the basic formulas used in our study.
Our task includes three numerical calculations: (i)$n(E)$,
(ii)  $ \displaystyle \bra \det \Delta(\mu)^{N_f/4} \ket_{E} $ and
(iii)  $ \displaystyle \bra O\det \Delta(\mu)^{N_f/4} \ket_{E} $.
Luo \cite{LUO} argued that if one stores the eigenvalues of the fermion matrix for all configurations,
then one can evaluate $\displaystyle \bra \det \Delta(\mu)^{N_f/4} \ket_{E} $
for any quark mass and flavor.
Let $\lambda_i(\mu)$ be the i-th eigenvalue of the massless fermion matrix
$\Delta(m_q=0)$ on a configuration with $E$. 
Then we obtain
\be
\bra \det \Delta(\mu)^{N_f/4} \ket_{E} 
= \bra ( \prod_i^{N_c V}(\lambda_i(\mu)+m_q))^{N_f/4} \ket_E.
\label{det}
\ee
Since (iii)  contains $O$, in general it is not calculable for 
any quark mass and flavor. However the chiral condensate $\bra\bar{\psi}\psi\ket$
which is obtained as the trace of the inverse fermion matrix 
can also be  given with the eigenvalues.
Namely, for $\bra\bar{\psi}\psi\ket$ 
we obtain
\be
 \bra (\frac1{V}Tr \Delta(\mu)^{-1}) \det \Delta(\mu)^{N_f/4} \ket_{E} 
=\bra \frac1{V}\sum_i \frac1{\lambda_i(\mu)+m_q} 
 ( \prod_i^{N_c V}(\lambda_i(\mu)+m_q))^{N_f/4} \ket_{E}.
\label{psidet}
\ee

\subsection{$g(U) \neq 1$}
Depending on the purpose, one may take $g(U) \neq 1$.
In Ref.\cite{Gocksch} the DOS for the complex phase of the fermion determinant
was calculated.
The DOS was defined as
\be 
n(E) =\int [dU] g(U) \delta(E-\theta(U)),
\ee
where $g(U)$ is set to $\displaystyle |\det \Delta|  e^{-\beta S_g(U)}$,
and $\theta(U)$ is the complex phase of the determinant. 
Then the expectation value of $O$ is given by
\be
\bra O \ket =\int^\pi_{-\pi} dE \bra O \ket_E n(E)e^{iE}/\int^\pi_{-\pi} dE  n(E)e^{iE}.
\ee
In Ref.\cite{RANDOM} the DOS for the number density was used and 
$g(U)$ is also set to $\displaystyle |\det \Delta|  e^{-\beta S_g(U)}$.

In these definitions parameters $\beta$ and $m_q$ are absorbed in $n(E)$ 
and we cannot vary $\beta$ and $m_q$.
Therefore, in this study we do not use these definitions with $g(U)\ne 1$.

\section{Simulations}

For the case of $N_f=2$, 
we have two chemical potentials $\mu_u$ and $\mu_d$ 
associated with the $u$ and $d$ quarks respectively.
If we introduce the isospin chemical potential $\mu_I$ as $\mu_I\equiv \mu_u=-\mu_d$, we obtain
\be
\displaystyle  [\det \Delta(\mu_u)\Delta(\mu_d)]^{1/4}
=[\det \Delta(\mu_I)\Delta(-\mu_I)]^{1/4}=|\det \Delta(\mu_I)|^{1/2},
\ee 
which is positive in the Monte Carlo measure. 
Therefore, for this $\mu_I$ the sign problem is absent and one can perform Monte Carlo simulations.
The natural generalization of the isospin chemical potential to $N_f \neq 2$ 
is that one takes 
$\displaystyle \det \Delta(\mu_I)^{N_f/4} \equiv |\det \Delta(\mu_I)|^{N_f/4}$.

\begin{figure}[htb]
\vspace{5mm}
\begin{center}
\epsfig{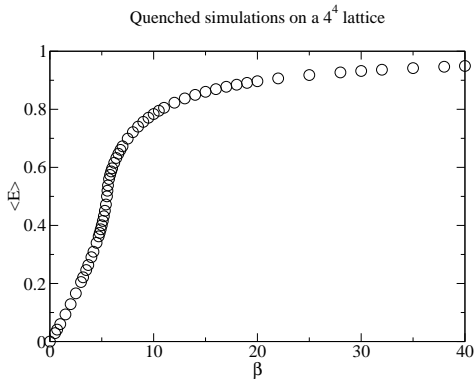}
\caption{\label{fig:plaqave}
Plaquette energy as a function of $\beta$.
}
\end{center}
\end{figure}

\begin{figure}[htb]
\vspace{5mm}
\begin{center}
\epsfig{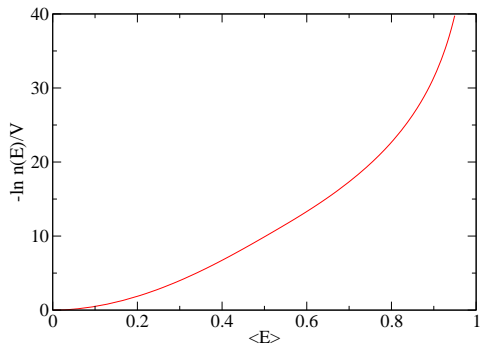}
\caption{\label{fig:integrated}
$-\ln n(E)/V$ as a function of plaquette energy $E$.
}
\end{center}
\end{figure}

We follow the implementation developed in Ref.\cite{LUO}.
The DOS $n(E)$ in eq.(\ref{dosn}) can be obtained using the quenched data as 
\be 
- \frac{\ln n(E)}{V} = 6 \int^E_0 dE^\prime \beta(E^\prime) + const.
\ee
First we make  quenched simulations on a $4^4$ lattice and determine the coupling constant $\beta(E)$ as 
a function of the plaquette $E$ ( Figure \ref{fig:plaqave} ). 
Then we integrate $\beta(E)$, numerically interpolating the data according to the trapezoidal rule. 
Figure \ref{fig:integrated} shows the result of $\displaystyle -\ln n(E)/V$.

\begin{figure}[hbt]
\vspace{5mm}
\begin{center}
\epsfig{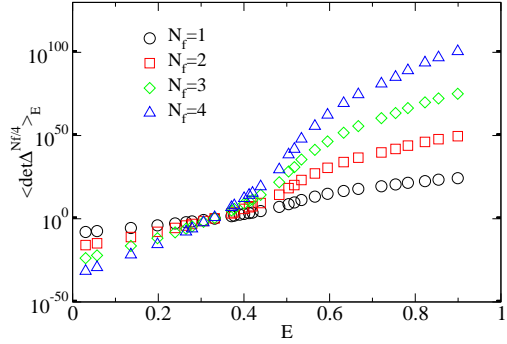}
\caption{\label{fig:detnfall}
Microcanonical average of $\det \Delta(\mu_I)^{N_f/4}$ at $\mu_I=0.2$ and $m_q=0.025$ as a function of $E$. 
}
\end{center}
\end{figure}

The most time consuming part of our method is 
the calculations of the microcanonical averages  $ \displaystyle \bra \det \Delta(\mu_I)^{N_f/4} \ket_{E} $ and 
 $ \displaystyle \bra O\det \Delta(\mu_I)^{N_f/4} \ket_{E}$ 
as a function of $E$, which contain the eigenvalue calculations.
In order to generate configurations at $E$ we use the over-relaxation method\cite{OVER,LUO}.
Starting from a configuration with $E$, we have made the over-relaxation update 
and saved 100 configurations at each $E$. 
Each configuration is separated by 100 over-relaxation updates.
About 30 values of $E$ are chosen in $E\in[0.05,0.95]$.
For each configuration we calculate the eigenvalues of the fermion matrix and store them.
Using these eigenvalues we can evaluate  $ \displaystyle \bra \det \Delta(\mu_I)^{N_f/4} \ket_{E} $ 
as in eq.(\ref{det}) for any quark mass and flavor.
Examples of $ \displaystyle \bra \det \Delta(\mu_I)^{N_f/4} \ket_{E} $  are shown in Figure \ref{fig:detnfall}.
Eq.(\ref{psidet}) is evaluated similarly. 
These data are interpolated by polynomials when we perform eqs.(\ref{dosn})-(\ref{dosn3}).

\section{Results}

\begin{figure}[hbt]
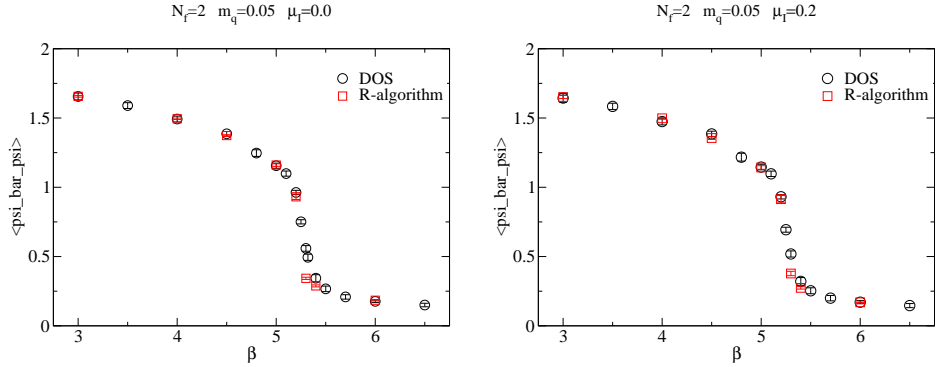

%\vspace{5mm}
\begin{center}
\epsfig{file=fig4left.eps,height=4.8cm}
\hspace{0.3cm}
\epsfig{file=fig4right.eps,height=4.8cm}
\caption{\label{fig:psinf2}
$\bra \bar{\psi}\psi\ket$ for $N_f=2$ at $m_q=0.05$ and at 
$\mu_I=0.0$(left) and  0.2(right).
Circles are from the DOS method.
The results from the R-algorithm are shown with squares.
}
\end{center}
\end{figure}

\begin{figure}[hbt]
%\vspace{5mm}
\begin{center}
\epsfig{file=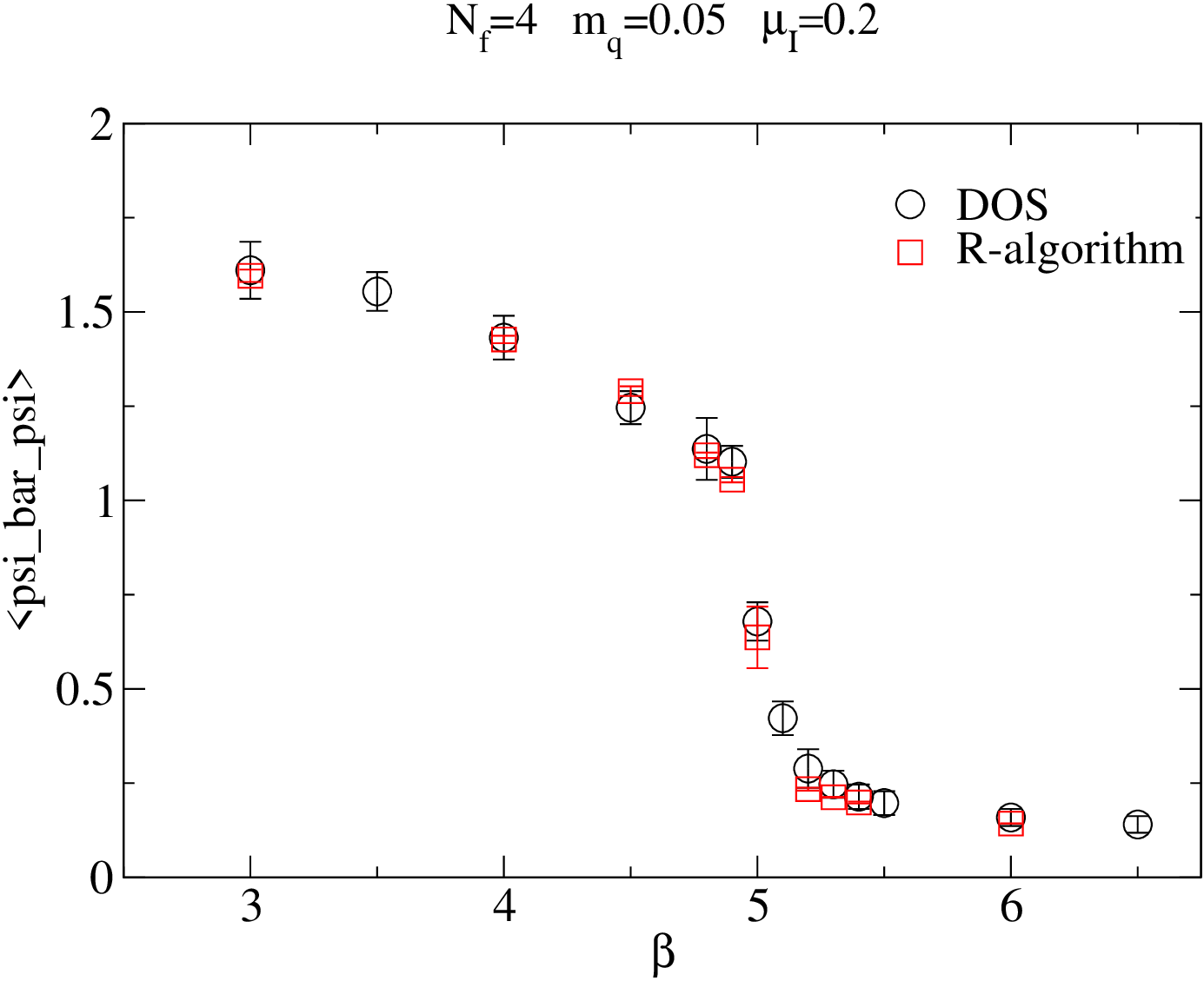,height=4.8cm}
\hspace{0.3cm}
\epsfig{file=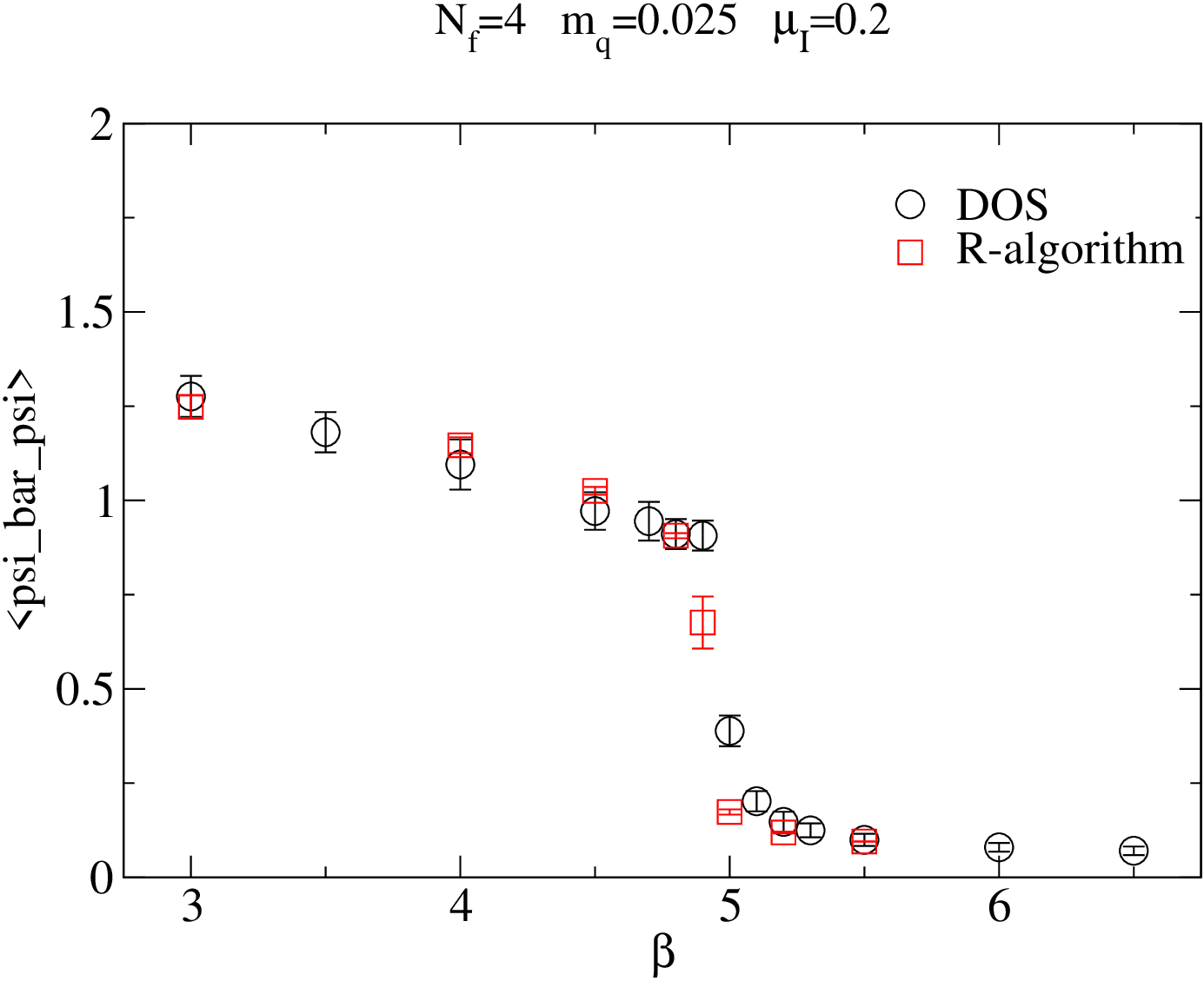,height=4.8cm}
\caption{\label{fig:psinf4mu02}
$\bra \bar{\psi}\psi\ket$ for $N_f=4$ at $\mu_I=0.2$ and 
at $m_q=0.05$(left) and 0.025(right).
}
\end{center}
\end{figure}

Figures \ref{fig:psinf2} and \ref{fig:psinf4mu02} compare results of $\bra \bar{\psi}\psi\ket$
between the DOS method and the R-algorithm\cite{R-algo}.
We see a good agreement between them in a wide range of $\beta$. 
A small  difference is seen in  
the phase transition region where 
the  $\bra \bar{\psi}\psi\ket$ changes rapidly.
In such a region we probably need careful analysis.

\begin{figure}[hbt]
\vspace{5mm}
\begin{center}
\epsfig{file=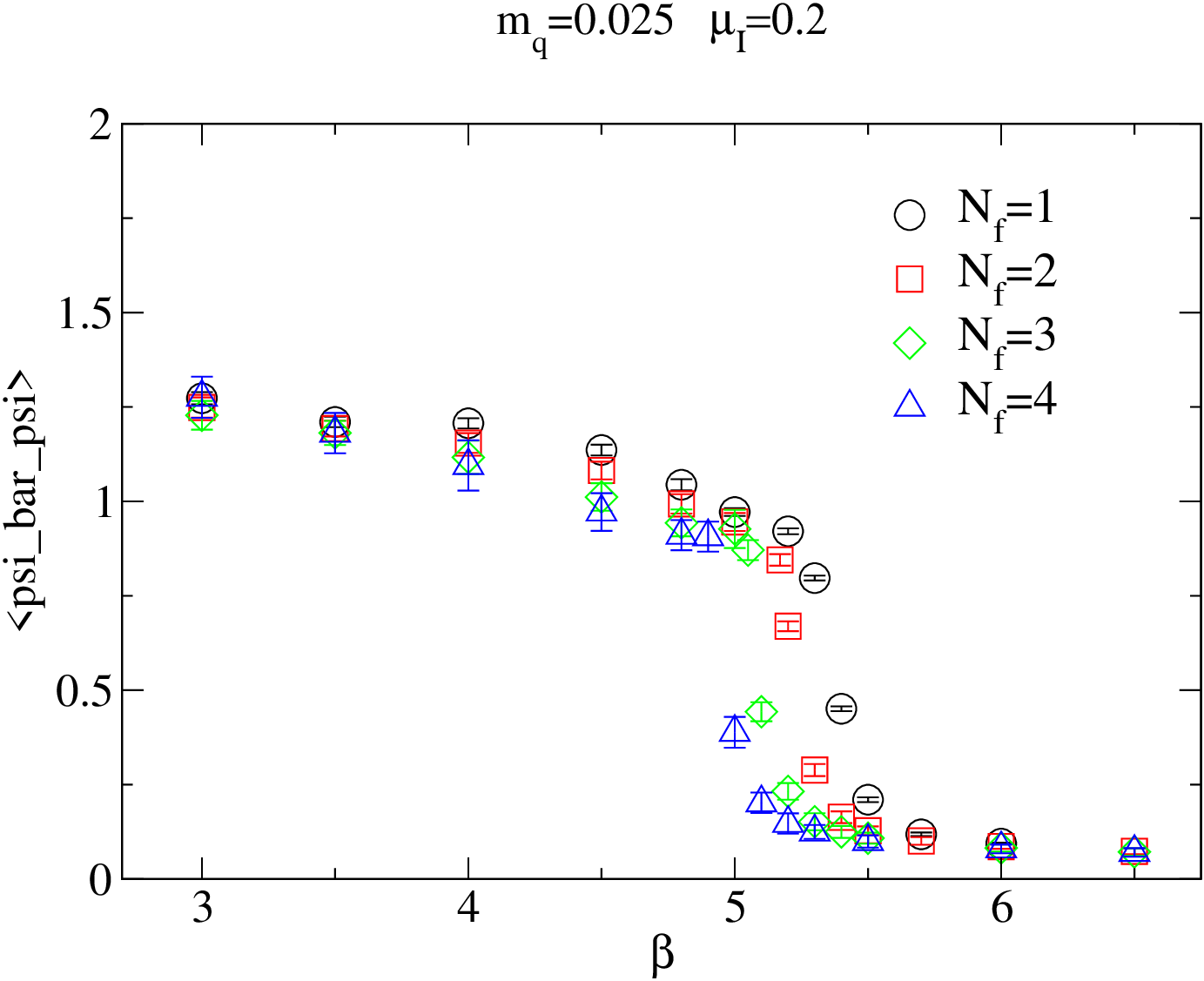,height=4.8cm}
\hspace{0.3cm}
\epsfig{file=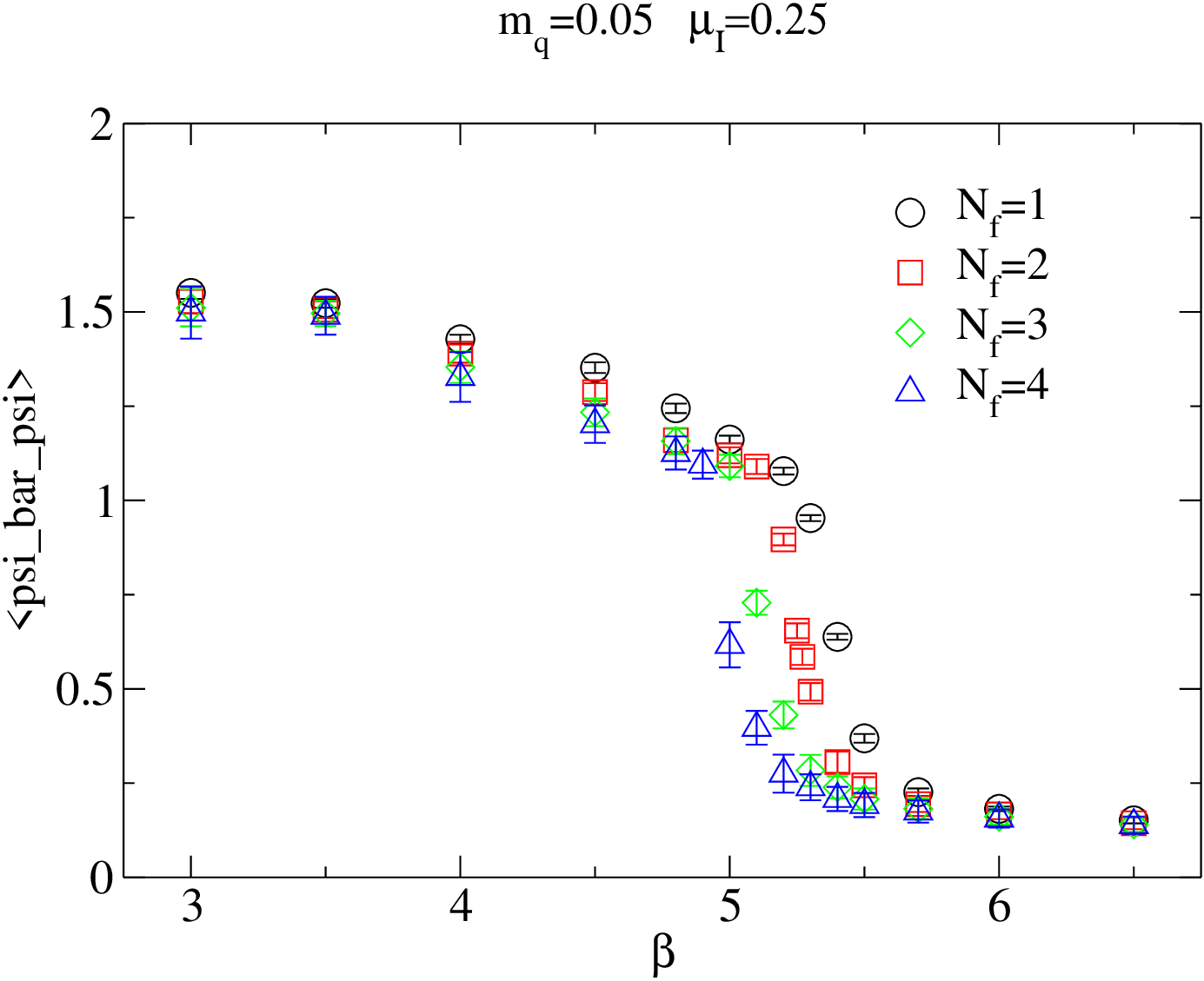,height=4.8cm}
\caption{\label{fig:psimu02nfall}
$\bra \bar{\psi}\psi\ket$ for different $N_f$.
(left) $m_q=0.025$ and $\mu_I=0.2$. (right) $m_q=0.05$ and $\mu_I=0.25$.
}
\end{center}
\end{figure}

\begin{figure}[hbt]
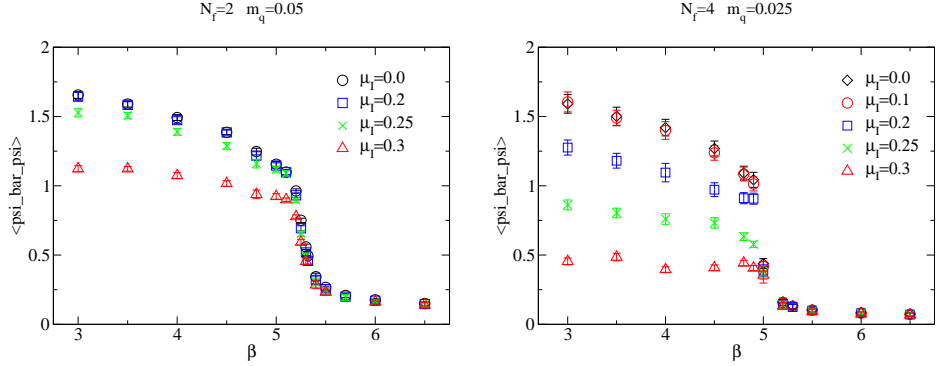

\vspace{5mm}
\begin{center}
\epsfig{file=fig7left.eps ,height=4.8cm}
\hspace{0.3cm}
\epsfig{file=fig7right.eps ,height=4.8cm}
\caption{\label{fig:psimuallmq0025}
$\bra \bar{\psi}\psi\ket$ for different $\mu_I$.
(left) $N_f=2$ and $m_q=0.05$. (right) $N_f=4$ and $m_q=0.025$. 
}
\end{center}
\end{figure}

Figure \ref{fig:psimu02nfall} shows $\bra \bar{\psi}\psi\ket$ for different $N_f$
at $m_q=0.025$ and $\mu_I=0.2$(left), and at $m_q=0.05$ and $\mu_I=0.25$(right) as  functions of $\beta$.
One can see that the critical coupling where the phase transition occurs
decreases as $N_f$ increases. 
Figure  \ref{fig:psimuallmq0025} shows  $\bra \bar{\psi}\psi\ket$ for various $\mu_I$. 
In the low-temperature phase $\bra \bar{\psi}\psi\ket$ decreases as $\mu_I$ increases.
On the other hand, in the high-temperature phase 
no visible difference can be seen among $\bra \bar{\psi}\psi\ket$.
Since the DOS method can explore various parameter space easily 
it is considered to be useful for exploring  a wide parameter space and for seeing a rough phase diagram.

In the DOS method we can take various combinations of parameters.
Let us consider the case of $N_f=1+1$ with non-degenerate quark masses $m_1$ and $m_2$.
In this case we must calculate the following microcanonical averages: 
\be
\displaystyle \bra |\det \Delta(m_1)|^{N_f/4} |\det \Delta(m_2)|^{N_f/4} \ket_{E},
\ee
\be
\displaystyle \bra  \bar{\psi}\psi(m_{i=1,2}) |\det \Delta(m_1)|^{N_f/4} |\det \Delta(m_2)|^{N_f/4} \ket_{E}.
\ee
Since the eigenvalues are stored, it is easy to calculate these microcanonical averages.
On the other hand, in the conventional algorithm such as R-algorithm one needs 
a differently implemented program to simulate $N_f=1+1$.
Such a program can be implemented but may become intricate.

\begin{figure}
\begin{center}
\epsfig{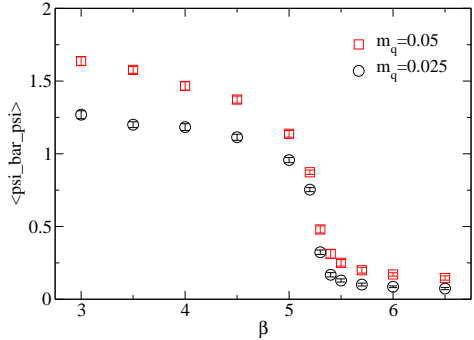}
\caption{\label{fig:nf1+nf1psimu02all}
$\bra \bar{\psi}\psi(m_q)\ket$ for $N_f=1+1$ at $\mu_I=0.2$ with different quark masses,
$m_q=0.05$ and $0.025$.
}
\end{center}
\end{figure}

\begin{figure}[hbt]
\vspace{5mm}
\begin{center}
\epsfig{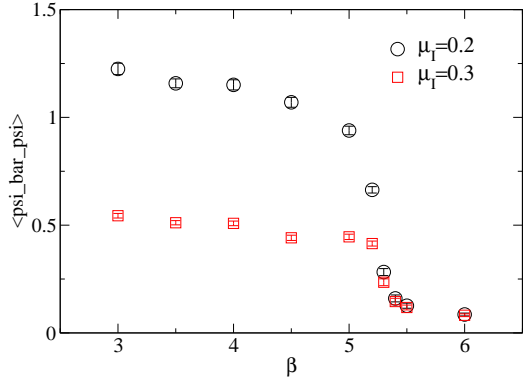}
\caption{\label{fig:mu0.2+mu0.3psiall}
$\bra \bar{\psi}\psi(\mu_I)\ket$ for $N_f=1+1$ at $m_q=0.025$ 
with different $\mu_I=0.2$ and 0.3  as a function of $\beta$.
}
\end{center}
\end{figure}

Figure \ref{fig:nf1+nf1psimu02all} shows $\bra \bar{\psi}\psi\ket$ for $N_f=1+1$ 
with different quark masses (  $m_1=0.05$ and $m_2=0.025$ ) and at $\mu_I=0.2$.
Since the two quarks have different masses, we plot two $\bra \bar{\psi}\psi(m_q)\ket$  
for $m_1$ and $m_2$.
Similarly, we can also consider non-degenerate isospin chemical potentials ( $\mu_1$ and $\mu_2$ ).
In this case we calculate
\be
\displaystyle \bra |\det \Delta(\mu_1)|^{N_f/4} |\det \Delta(\mu_2)|^{N_f/4} \ket_{E},
\ee
\be
\displaystyle \bra  \bar{\psi}\psi(\mu_{i=1,2}) |\det \Delta(\mu_1)|^{N_f/4} |\det \Delta(\mu_2)|^{N_f/4} \ket_{E}.
\ee
Figure \ref{fig:mu0.2+mu0.3psiall} shows $\bra \bar{\psi}\psi\ket$ for $N_f=1+1$ 
with different chemical potentials ( $\mu_1=0.2$ and $\mu_2=0.3$ ) at $m_q=0.025$.

\begin{figure}[hbt]
\vspace{1cm}
\begin{center}
\epsfig{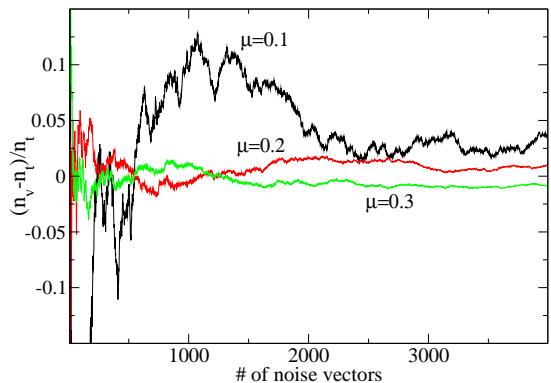}
\caption{\label{fig:convtest.eps}
Relative error of the number density for different $\mu$ at $m_q=0.025$
as a function of the number of noise vectors.
A representative quenched configuration at $\beta=5.5$ was used.
$n_v$ is the number density calculated by the noise method and $n_t$ is the true value of the number density.
$Z_2$ noise vectors are used for these calculations.
We also used Gaussian noise vectors and obtained similar results.
}
\end{center}
\end{figure}

\begin{figure}[hbt]
\vspace{1cm}
\begin{center}
\epsfig{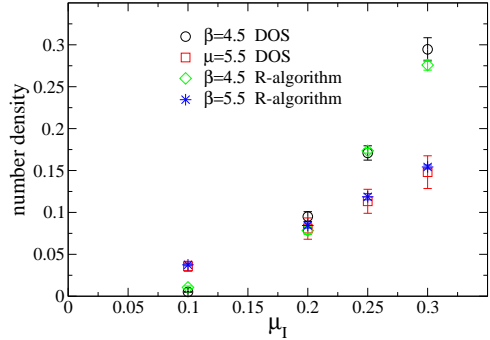}
\caption{\label{fig:number}
Number density as a function of $\mu_I$.
}
\end{center}
\end{figure}

Next we show the results of the number density $n_d$ defined by
\be
n_d=\frac1V\frac{\ln Z}{\partial\mu}=
\bra\frac1V Tr\frac1{ \Delta(\mu)}\frac{\partial \Delta(\mu)}{\partial\mu}\ket.
\label{eq:number}
\ee
Since this  can not be evaluated with  the eigenvalues of $\Delta(\mu)$ only,
in order to obtain $\displaystyle \bra O|\det \Delta(\mu)|^{N_f/4} \ket_{E}$
for the number density
we must  calculate the number density on each configuration.

Usually Eq.(\ref{eq:number}) is evaluated by the noise method.
Using noise vectors $R_i$ having the property $\bra R_i^\dagger R_j\ket=\delta_{ij}$, 
the trace calculation can be replaced by
\be
 Tr\frac1{ \Delta(\mu)}\frac{\partial \Delta(\mu)}{\partial\mu}
=\frac1N \sum_{i=1}^N R_i^\dagger \frac1{ \Delta(\mu)}\frac{\partial \Delta(\mu)}{\partial\mu}R_i,
\ee
where $N$ is the number of noise vectors.
Although this is true for $N \rightarrow \infty$, 
we find that the convergence of the noise method is extremely slow on a $4^4$ lattice. 
A similar result was reported in Ref.\cite{FKT}.
Figure \ref{fig:convtest.eps} shows the convergence measured as relative error
as a function of the number of $Z_2$ noise vectors\cite{Z2}.  
Typically,  $O(1000)$ noise vectors are needed to have a reasonable value. 
In this analysis, instead of using the noise method we calculated the number density exactly by
calculating each diagonal element of   
$\displaystyle \frac1{ \Delta(\mu)}\frac{\partial \Delta(\mu)}{\partial\mu}$.
In general such  calculations are computationally costly 
and should be avoided. However, our lattice size is sufficiently small to perform the exact calculation.
Thus, here we have adopted the exact calculation.

Figure \ref{fig:number} shows the number density as a function of $\mu_I$.
The results from the R-algorithm are also plotted in the figure.
The two results are in good agreement.

\section{Summary and outlook}

We have given the general formulas of the DOS method including dynamical
fermions.
The case of "$g(U)=1$" corresponds to that used by Luo\cite{LUO}.
Based on the implementation by Luo,
we have calculated  $\bra \bar{\psi}\psi\ket$ and the number density 
at finite isospin densities by the DOS method
and made a comparison with results from the R-algorithm. 
The two results were found to be in good agreement.
The DOS method can be applied for various combinations of parameters.
We have calculated $\bra \bar{\psi}\psi\ket$ for various values of $N_f$ and $\mu_I$,
and also for non-degenerate quark masses and for different isospin chemical potentials. 
Especially it is emphasized that 
for non-degenerate quark masses and for different isospin chemical potentials
it is not easy to perform Monte Carlo simulations by the conventional algorithm 
as the R-algorithm but these calculations are easily performed in the DOS method
by keeping all eigenvalues of the fermion matrix. 

The limitation of the DOS method may appear on a large lattice.
The measurement of the DOS method is done 
in the microcanonical ensemble and $\bra \det \Delta \ket$ is
treated as an observable. 
Since the microcanonical and full ensembles are very different, $\bra \det \Delta\ket$ is  
expected to fluctuate largely as the volume of the system increases, 
which may limit the available lattice to a small one.   

In principle, the DOS method can be applied for QCD with a baryon chemical potential.
However, there still remains the sign problem.
For example, for $N_f=4$ we have $\displaystyle \bra \det \Delta(\mu_B) \ket_{E} =\bra |\det \Delta(\mu_B)|e^{i\theta} \ket_{E}$
and if $e^{i\theta}$ fluctuates significantly, which is expected to occur for larger $\mu_B$, 
one cannot obtain meaningful values for the microcanonical average.
We attempted to calculate $\bra |\det \Delta(\mu_B)|e^{i\theta} \ket_{E}$ but
for $\mu_B >0.2$ 
we could not obtain statistically meaningful values\footnote{Here $\mu_B$ is defined as the quark chemical potential.}.
On the other hand, for $\mu_B< 0.2$ the results were stable but 
there is no visible difference between 
the results of $\bra \bar{\psi}\psi\ket$ at $\mu_I$ and at $\mu_B$;  
for $\mu_B<0.2$ the effect of the complex phase is small and thus we do not see
any difference. This is consistent with the results of Refs.\cite{TOUSSAINT,SNT}.

In this study for each value of $\mu$ we calculated the eigenvalues.  
However, if we use a matrix reduced to two time slices\cite{REDUCED,REVIEW3},
we can calculate the determinant at any $\mu$.
The determinant for this case is expressed as
\be
\det \Delta(\mu)=\det(P+e^{N_t \mu})\times e^{3V\mu},
\ee
where $P$ is a $\mu$-independent matrix and $N_t$ is the number of lattice sites in the time direction.
Keeping the eigenvalues of $P$, we can calculate the determinant at any $\mu$.
In this case, however, the tradeoff is that the determinant is not calculable at any $m_q$.
Thus, if one wishes to obtain the behavior with $\mu$ at fixed $m_q$ 
this reduction method may be useful.

\section*{Acknowledgments}
These calculations were carried out on the SX-5 at RCNP, Osaka University.
The author would like to thank Ph. de Forcrand, X.Q.Luo and A.Nakamura 
for helpful comments and discussions.

\end{document}